\documentclass[prd,superscriptaddress,preprintnumbers,amsmath,amssymb,nofootinbib,tightenlines,11pt]{revtex4}
\usepackage[latin1]{inputenc}
\usepackage{graphicx}
\usepackage{color}
\usepackage{multirow}
\usepackage{amsmath,amssymb}
\usepackage{hyperref}
\usepackage{url}
\usepackage{slashed}
\usepackage{subfigure}
\usepackage[usenames,dvipsnames]{xcolor}
\newcommand{\be}{\begin{equation}}
\newcommand{\ee}{\end{equation}}
\newcommand{\bea}{\begin{eqnarray}}
\newcommand{\eea}{\end{eqnarray}}

\newcommand{\vx}{\vec x}
 
\newcommand{\cdp}{d_p} 
\newcommand{\dd}{\delta}

\newcommand{\dm}{D} 
\newcommand{\dmeff}{{\dm_{\mathrm{eff}}}}
 
\newcommand{\past}{\mathrm{Past}}
\newcommand{\fut}{\mathrm{Future}}
\newcommand{\cdpp}{\mathbf{d}_p}

\begin{document}

\title{Echoes of Asymptotic Silence in Causal Set Quantum Gravity}

\author{Astrid Eichhorn}
  \email{a.eichhorn@thphys.uni-heidelberg.de}
\affiliation{Institut f\"ur Theoretische
  Physik, Universit\"at Heidelberg, Philosophenweg 16, 69120
  Heidelberg, Germany}
\author{Sebastian Mizera}
\affiliation{Perimeter Institute for Theoretical Physics, Waterloo, Ontario N2L 2Y5, Canada}
\affiliation{Department of Physics \& Astronomy, University of Waterloo, Waterloo, Ontario N2L 3G1, Canada}
\author{Sumati Surya}
\affiliation{Raman Research Institute, C.V. Raman Avenue, Sadashivnagar, Bangalore
560 080, India}

\begin{abstract}
We explore the idea of asymptotic silence in causal set theory and find that causal sets
 approximated by continuum spacetimes exhibit  behavior 
akin to asymptotic silence. 
 We make use of an intrinsic definition of spatial distance between causal set elements in the discrete analogue of a spatial hypersurface. Using numerical simulations for causal sets approximated by $\dm=2,3$ and $4$ dimensional Minkowski spacetime, we show that while the discrete distance rapidly converges to the continuum distance at a scale roughly an order of magnitude larger than the  
  discreteness   scale, it is significantly larger on small scales. This allows us to define an  effective dimension 
 which exhibits dimensional reduction in the ultraviolet, while 
 monotonically increasing to the continuum dimension with  increasing 
 continuum distance. We interpret these findings as manifestations of asymptotic silence in causal set theory. 
 \end{abstract}

\maketitle

\section{Introduction}

Several approaches to quantum gravity exhibit an ultraviolet reduction of the spectral dimension from four to (close to) two dimensions, where the spectral dimension is defined via a Riemannian diffusion process. These models include Dynamical Triangulations \cite{Ambjorn:2005db,Benedetti:2009ge,Anderson:2011bj,Gorlich:2011ga,Cooperman:2014uha,Laiho:2016nlp,Laiho:2011ya}, asymptotically safe gravity \cite{Lauscher:2005qz,Lauscher:2005xz,Reuter:2011ah,Rechenberger:2012pm,Calcagni:2013vsa}, Ho\v{r}ava-Lifshitz gravity \cite{Horava:2009if,Sotiriou:2011mu}, and several other models \cite{Calcagni:2012rm,Amelino-Camelia:2013cfa,Arzano:2014jfa,Alkofer:2014raa,Calcagni:2014cza,Ronco:2016rtp,Arzano:2016fuy}. It was argued in \cite{Carlip:2009kf,Carlip:2009km,Carlip:2012md,Carlip:2011tt,Carlip:2016qrb} that this scale dependent dimension is related to the phenomenon of ``asymptotic silence'' in quantum spacetime. Asymptotic silence describes the ``narrowing" or sharp focusing of lightcones near the Planck scale, leading to a decoupling of seemingly nearby worldlines. 
Near the singularities of a classical spacetime for example, asymptotic silence 
leads to a Kasner-like behavior of the metric, which exhibits dimensional reduction \cite{Hu:1986exa,Heinzle:2007kv}.  Near the space-like singularity communication between different worldlines is prohibited. 
 
In a quantum spacetime, properties such as the dimensionality can become scale-dependent. In particular, 
 appropriately defined dimensional estimators can change as one ``zooms in" to smaller scales, while they converge to their continuum 
values at larger scales. 
 From a more fundamental point of view, one might even expect the standard manifold structure of spacetime to break down in the quantum regime.

Since the diffusion process that determines the spectral dimension in all the quantum gravity approaches
mentioned above is Riemannian, relating it to a Lorentzian 1+1 dimensional spacetime is far from
obvious. Thus the connection between dimensional reduction and asymptotic silence is an indirect one.
On the other hand,  it may be possible to probe the quantum properties of the causal structure more
directly in a manifestly Lorentzian approach like causal set theory (CST).

In CST the spacetime continuum is replaced by a fundamentally discrete 
structure, which is a locally finite partially ordered set. Spacetime discreteness is implemented by requiring that any 
spacetime region of finite volume in the continuum contains a finite number of fundamental spacetime   ``atoms" and that this number, on average, is the continuum spacetime volume with respect to the \emph{physical} discreteness scale.

In \cite{Eichhorn:2013ova} it was shown that results of numerical simulations of diffusion on causal
sets show an increase of the spectral dimension on short scales\footnote{In \cite{Belenchia:2015aia}, a nonlocal d'Alembertian was used to extract
a spectral dimension which 
showed an initial
dimensional increase towards the UV and 
a subsequent dimensional reduction at
even smaller scales.
Thus the results from numerical simulations \cite{Eichhorn:2013ova}
appear to be in contradiction with  
these analytical results. 
However the nonlocal d'Alembertian used in \cite{Belenchia:2015aia} 
is 
a fully continuum version of the causal set d'Alembertian.  
Its use in the far UV cannot therefore accurately reflect the effect of causal set discreteness. On the other hand, away from this scale the dimensional increase matches that of \cite{Eichhorn:2013ova}.
} 
which was interpreted as a possible signature of asymptotic silence in  
\cite{Carlip:2015mra}. The spectral dimension is obtained from a diffusion process in a fictitious, external time variable $\sigma$, and measures the return probability of the corresponding random walker to its starting point. A generalization
{ introduced}
in \cite{Eichhorn:2013ova} considers two causal random walkers with the spectral dimension given by 
\be
d_s(\sigma) = -2 \frac{d\ln P(\sigma)}{d\ln \sigma},  
\ee
where $P(\sigma)$ is their meeting probability after time $\sigma$. 
A decrease in the meeting probability would therefore lead to an increase in the spectral dimension. This is what one might expect to happen in an asymptotically silent spacetime in which the worldlines are further apart because of a narrowing of the lightcones.
Thus, dimensional increase could be attributed to asymptotic silence, as advocated in \cite{Carlip:2015mra}.

In this work we will explore this intriguing suggestion more directly. As  stated in the usual way, the notion
of asymptotic silence  requires a foliation of spacetime by spacelike hypersurfaces 
 since it refers to the decoupling of \emph{spatially} ``nearby''  points. On the other hand, causal set quantum gravity is an intrinsically
covariant or \emph{spacetime} approach to quantum gravity and the notion of a foliation  is 
ill
defined in a causal set.    It is therefore not immediately obvious
how to translate what it means for a light cone to ``narrow'' as a function of time in this setting.

Instead, an ``echo'' of asymptotic silence in causal set theory comes unexpectedly from a
calculation of causal set homology \cite{Major:2006hv,Major:2009cw}.  Starting from an inextendible
antichain $A$, which is the discrete analogue of a Cauchy hypersurface (for details on causal set concepts and terminology, see Sec.~\ref{secone}),  
 a family of homology groups was constructed as a function of $n \in \mathbb{N}$,  which converges to give the
continuum homology for large enough $n$, but not for smaller $n$. 
Here $n$ is a dimensionless measure of scale with small $n$ corresponding to the ultraviolet (UV) or  discreteness scale 
and large $n$ to the continuum. 
In particular, for small $n$ there are a  large number of spatially
disconnected regions or ``islands'' in the underlying causal set, even for a connected spacetime.  As $n$ increases, these islands join together and eventually
converge to a single connected region. The existence of these islands is  similar to the enhanced
separation between neighbouring worldlines in the UV which characterise asymptotic silence. Indeed, as 
we  will demonstrate, a similar construction can be used to define asymptotic silence in causal set theory. 

In Section \ref{secone} we begin with a brief review of CST elucidating  some of its key features  which distinguish it from other approaches to quantum gravity. In Section \ref{sectwo} we give a definition of spatial distance $\dd$ in the discrete analogue of a Cauchy hypersurface  and use it to show asymptotic silence in CST. We argue that $\dd$ is always greater than the (normalised) continuum proper distance $\cdpp$ between the elements  and that because of fluctuations arising from randomness, the
difference in $\dd$ and  $\cdpp$ grows as $\cdpp$ decreases.

This increased distance in the ultraviolet can be interpreted as 
a manifestation of asymptotic silence. We  present results from numerical simulations for causal sets that are approximated by  $\dm=2,3$ and $4$ Minkowski spacetime which strongly support this  argument.  Interestingly, we find that the approach to the continuum occurs at roughly the same length scale in all dimensions, $\sim 10 \times  l_c$, where $l_c$ is  the 
discreteness scale.  We use $\dd$ and $\cdpp$ to  
define an   ``effective dimension'' $\dmeff$ 
which exhibits dimensional reduction in the ultraviolet.

It is important to point out that 
the discreteness
scale $l_c$ 
defines what we refer to as the ultraviolet  scale 
for causal sets
in this context 
and is distinct from the deep quantum regime characterised by the Planck scale $l_p$. This distinction is important in CST since
the continuum need not play any role near $l_p$. 
Indeed, the 
 full quantum theory of causal sets must include those that are not manifoldlike. 
Manifoldlike properties would only emerge after coarse graining of the causal set, at  scales $l_c \gg l_p$.  The  phenomena we describe in this paper are therefore due to the {\it traces} of discreteness which are still present in the continuum approximation, at scales  $\sim l_c$.  
Indeed, in this sense causal sets are quantum even without quantum interference, since the causal set underlying a given continuum spacetime carries a discreteness scale.  Thus, causal sets differ from the continuum already at the {\it kinematical} level.  The effect of this discreteness can lead to non-classical behaviour of test particles, as in the case of swerves \cite{Dowker:2003hb,Philpott:2008vd} or the scalar field Green's function \cite{Sorkin:2007qi}. From this perspective it is therefore reasonable to ask if causal sets exhibit traces of asymptotic silence even without probing the deep quantum regime.

\section{An overview of CST} 
\label{secone} 

The causal-set approach to quantum gravity  is built  on two fundamental principles: causality and discreteness  \cite{Bombelli:1987aa} (for reviews see, e.g.,
\cite{Sorkin:2003bx,Dowker:2005tz,Henson:2010aq,Dowker:aza}). The former is motivated by powerful theorems in Lorentzian geometry which state that under very weak causality conditions the causal structure poset gives the conformal geometry of the spacetime and also its topology and dimension \cite{Malament, Hawking:1976fe}. The motivation for the latter comes from a variety of sources, such as the suggestion for the existence of a
cut-off in the 
description
of black hole entropy.  
This leads to a minimalistic  mathematical structure for quantum spacetime, which is that of a locally finite partially ordered set or causal set $C$
defined as follows. 

 A causal set $C$ is a countable collection of elements which are partially ordered via a relation $ \prec $ which is acyclic (${\rm if}\; x \prec y\; {\rm \, and}\; y \prec z\,  \Rightarrow x=y$),  transitive, $\forall x,y,z \in C: x \prec y,\; y \prec z\,  \Rightarrow  x \prec z$, and locally finite ($\forall x,y \in C: {\rm card}(\{z \in C: x \prec z \prec y\}) < \infty$).

Importantly, the requirement of  local finiteness ensures that finite spacetime volumes in the continuum approximation contain a finite number of causal set elements, thus ensuring a fundamental discreteness. 
The continuum spacetime arises as an approximation of the underlying causal set with the order relation $\prec$ corresponding to the continuum causal order  and such that the number of elements in a given region corresponds to the spacetime volume in units of a fundamental minimum volume. The approximation is implemented via a Poisson process (see \cite{Henson:2010aq,Dowker:2005tz,Sorkin:2003bx,Dowker:aza})  so that the number of elements is equal to the spacetime volume on average. 

It is this randomness that gives rise to 
interesting  phenomenology \cite{Philpott:2008vd,Belenchia:2014fda,Saravani:2015rva,Belenchia:2015ake,
  lambdapaper, Sorkin:1997gi, everpresent, Barrow:2006vy, Zwane:2017,Afshordi:2012ez,Sorkin:2016pbz}. 
In particular it ensures that despite discreteness, Lorentz invariance is not violated \cite{LLI,Glaser:2013pca}, but instead leads to non-locality.

CST differs crucially  from other discrete approaches to quantum gravity in that the discreteness is assumed to be physical, 
rather than a mathematical tool for regularising the theory. For instance in (Causal) Dynamical Triangulations \cite{Ambjorn:1998xu,Ambjorn:2006jf,Ambjorn:2011cg,Ambjorn:2012jv}, physics 
can only emerge in the continuum limit since 
the discretisation is introduced as an unphysical  regulator. In CST,  on the contrary,  since the  
discreteness  
is physical, 
the continuum limit is 
not. Thus one refers to the continuum \textit{approximation} rather than the continuum  {limit}.

Causal sets come in two distinct types: Those which are approximated by  $\dm$ dimensional spacetimes, and those which are not. 
The latter dominate the set of all causal sets entropically \cite{KR}, and hence an important question in CST is whether a suitable dynamics can be found that will suppress the entropy in favour of manifoldlike causal sets in the appropriate limit (see \cite{Rideout:1999ub, Benincasa:2010ac,Benincasa:2010as,Dowker:2013vba, Surya:2011du, Henson:2015fha} for more discussion on causal set dynamics). 
Since CST is a \textit{quantum} theory of {\textit{all}} causal sets,   one does not expect those that are manifoldlike to play a significant role in the deep quantum regime; the continuum approximation becomes relevant only  at larger scales $l_c \gg l_p$.
Nevertheless, the remnant of the quantum nature of the causal set manifests itself near the discreteness scale $\sim l_c$, 
and it is  this region that we explore\footnote{ Because of Lorentz invariance it is important to remember that the discreteness scale is characterised by the spacetime volume $V_c=l_c^{\dm}$ rather than a length scale.}.

\section{Asymptotic Silence in CST} 
\label{sectwo}

An \textit{antichain} in $C$ is a set of unrelated elements and is the analogue of a set of spacelike related elements in a spacetime.  The analogue of a Cauchy hypersurface in a globally hyperbolic spacetime is an
\textit{inextendible antichain}  $A\subset C$, which is an antichain $A$ such that all elements in $C$ that are not in $A$  belong to its past or future, where 
 for any set $S \subset C$
\be 
\fut(S) \equiv \{ e\in C| \exists \, s \in S,   s\prec e \},   \quad \past(S) \equiv \{ e\in C| \exists \, s \in S,  s \succ e \}.   
\ee
Inextendibility of $A$ thus ensures  that any other element in $C$ is  either to the past or the future of an element in $A$, so that $C$ can be divided  into the non-overlapping regions $C=A \sqcup \fut(A) \sqcup \past(A)$.

\begin{figure}[!t]
 \includegraphics[width=0.8\linewidth,clip=true,trim=10cm 10cm 2cm 8cm]{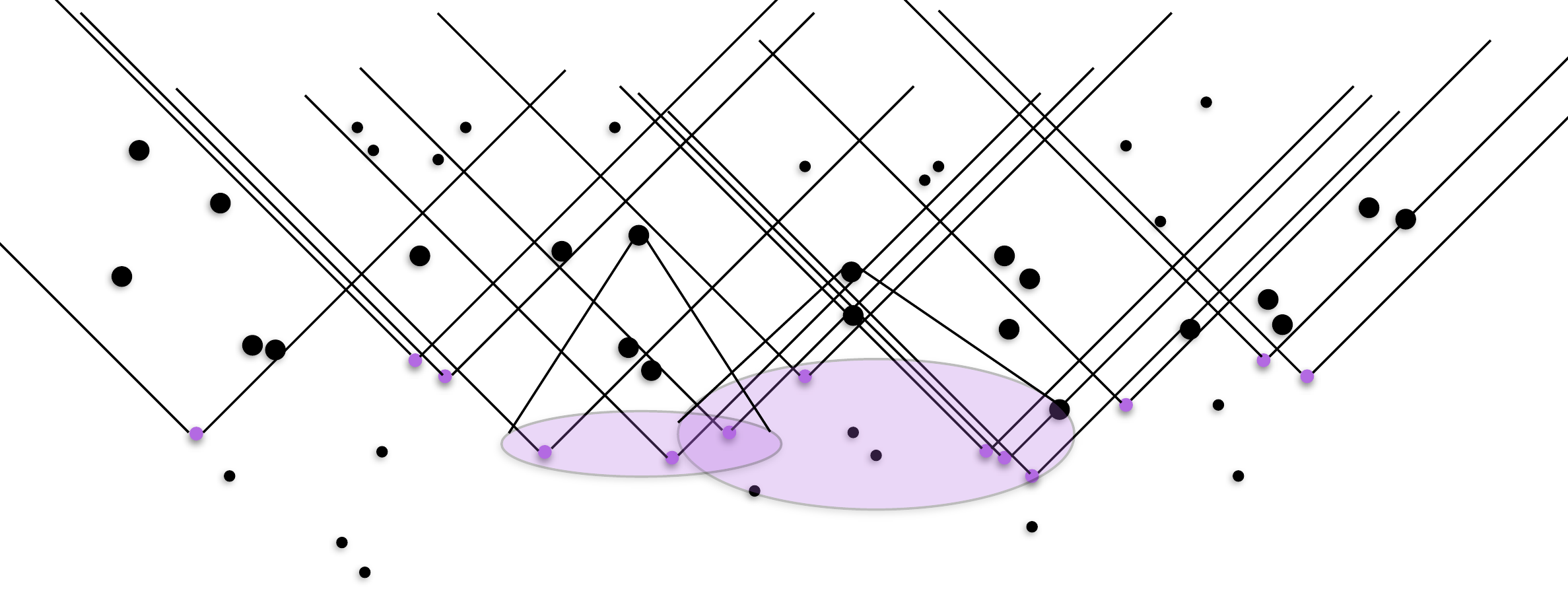}
 \caption{\label{fig:illustration_antichain}  We sketch an inextendible antichain (purple dots) in a sprinkling into two-dimensional Minkowski spacetime. Elements in a thickening with $v=5$ are shown as fat dots, and selected past and future lightcones are indicated.
}
 \end{figure}

$A$ itself contains no information, except its cardinality. Nevertheless, as explored in \cite{Major:2006hv,Major:2009cw}
it can borrow structure from $\fut(A)\subset C$ by ``evolving'' $A$  to a  future neighbourhood. For every  $e\in \fut(A)$, the cardinality 
\be
v(e) ={\rm card} \left(\past(e)\cap (A\cup \fut(A))\right)
\ee
 can be used to foliate $\fut(A)$ by constant $v$ ``slices'',  cf.~Fig.~\ref{fig:illustration_antichain}.
 Though these slices are themselves antichains, they need not be inextendible, but this will not matter to the analysis. 
A \textit{thickened antichain} is defined as  
\begin{equation} 
T_v(A) \equiv \{e \in \fut(A)\,|\, v(e)\leq v \}. 
\end{equation} 
$T_v(A)$ has more structure than $A$ itself and hence the potential to encode continuum geometrical and topological information for large enough $v$, see \cite{Major:2005fy}.  Conversely, for small $v$, one expects the continuum approximation to break  down. 

In \cite{Major:2006hv,Major:2009cw} an order theoretic construction capturing the spatial homology of a manifold-like causal set was given.  The homology was seen to be a function of $v$ but stabilised after a certain critical value.  The continuum topology is therefore emergent at large $v$, whereas discreteness manifests itself at small $v \sim V_c$.     For small $v$ the simplicial complex has several disconnected ``islands'' leading to a large zeroth betti number even when the continuum is connected.  As $v$ increases, these islands merge into a single connected component.

Finding a connection between stable homology and asymptotic silence is not obvious, but it provides an important clue. In particular, the discrete homology for a given $v$ is constructed from the causal structure of the elements in the thickening to the original antichain. This suggests a definition of a spatial distance on the inextendible antichain. For small $v$ the disconnected islands suggest a decoupling of elements in the causal set that are nevertheless spatially close in the continuum. Exploiting this connection gives us a useful definition of asymptotic silence compatible with causal set discreteness.

We look to the continuum for guidance on how to proceed.  In $\dm$-dimensional Minkowski spacetime,
consider the events $p=(0,\vx_1)$ and  $q=(0,\vx_2)$. Their proper distance $\cdp=|\vx_1-\vx_2|$ can be related
to a spacetime volume as follows. Without loss of generality, let the origin $(0,\vec{0})$ lie at the midpoint of  $p$ and
$q$ in the $t=0$ slice. The  chronological past of the event $r=(\cdp/2,\vec{0})$ does not include
$p,q$, but its closure (which in this case is equal to its causal past) does. The past volume of $r$ up to this  slice is
given by 
\begin{equation} 
\label{voldef} 
V=\zeta_\dm\, \cdp^\dm, 
\end{equation} 
where
\be\label{zeta}
\zeta_\dm= \frac{\pi^{(\dm-1)/2}}{2^\dm \dm\, \Gamma \left(\frac{\dm+1}{2} \right)}.
\ee
In Minkowski spacetime $\cdp$ is a distance function on the $t=0$ slice, since it naturally satisfies the triangle
inequality 
\begin{equation}
\cdp(p,q) \leq \cdp(p,s)+\cdp(q,s) 
\end{equation} 
for all $p,q,s$ in the $t=0$ hypersurface.  Indeed, this reproduces exactly the distance on the
$t=0$ slice.

Casting the distance into this form makes {it easier to construct} the causal set analogue. For a given pair of elements $p,q \in A$, let $r \succ p, q$ such that $\past(r)\ni p,q$. Let $n(r)=\past(r)\cap \fut(A)$ and let $n$ denote the infimum of $n(r)$ over all such $r$.  We then define the discrete causal set spatial distance as \begin{equation}\label{delta} \dd = \biggl(\frac{n}{\zeta_\dm}\biggr)^\frac{1}{\dm}.  \end{equation}
How are we to compare $\cdp$ which is dimensionful to the dimensionless $\dd$? For a causal set that is approximated by $\dm$-dimensional Minkowski spacetime it is natural to cast the proper distance in the continuum in units of the spacetime cut-off  $V_c$, 
\be
\cdpp \equiv \biggl(\frac{V}{\zeta_\dm \, V_c  }\biggr)^\frac{1}{\dm},  
\ee 
in order to facilitate the  comparison with $\dd$.

\begin{figure}[!t]
\begin{center}
\includegraphics[width=0.6\linewidth,clip=true,trim=2cm 10cm 3cm 9cm]{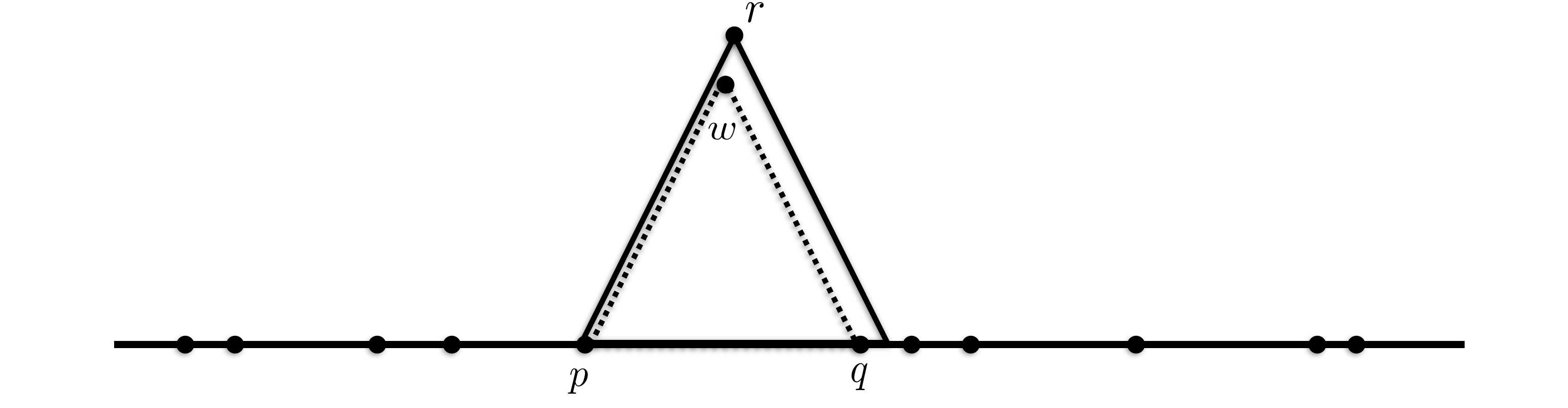}
\end{center}
\caption{\label{fig:illustration_AS}  An illustration showing why $\cdpp<\dd$ at small scales.  For two elements on the  $t=0$ slice, $\cdpp$ is obtained from the past volume of the  spacetime event $w$. In the  causal set the probability for an element  to lie on $w$ is  zero, and hence $\dd$ must be defined using the past of some $r\in C$ with $r \succ w$ which is larger.}
\end{figure}

If we make the unjustified assumption that the antichain  $A$ lies  exactly on the $t=0$ slice, we would find that $\dd$ is strictly  larger than $\cdpp$ since the probability of $r $ being to the causal but not the chronological future of both $p$ and $q$ is zero. In other words, since null surfaces are a set of measure zero the probability that a pair of elements in a causal set are null related   is zero. Thus  ${\rm Past}(r) \cap \fut(A)$ is always slightly larger than ${\rm Past}((d/2,\vec0)) \cap \fut (A)$ in the continuum (see Fig.~\ref{fig:illustration_AS} for an illustration).

Moreover, at small scales, the relative fluctuations in the number of elements in a given spacetime region is larger, and hence the overestimation is substantially larger.  As $n$ increases,   it is  reduced so that  $\dd$ 
converges to $\cdpp$.   In the continuum, 
$\cdpp$ translates 
into the shortest time taken for the worldlines of two observers starting at $p$ and $q$ to meet, with respect to the discreteness scale. Since $\dd > \cdpp$, so too is the ``time'' taken for them to meet in the underlying causal set.  Thus observers are further away from each other in the causal set than they would be in the continuum.  This then  is a direct manifestation of asymptotic silence.

Of course, the assumption that there exists  such a flat antichain $A$ is incorrect and the comparisons must be made more carefully. Given a sprinkling of $C$ into a spacetime $(M,g)$ with compact spatial hypersurfaces, consider any inextendible antichain $A \subset C$. The proper distance $\cdp$ of any pair $p,q \in A$ can be calculated from the embedding and expressed in terms of the cut-off $l_c$. However, its relationship to the volume $V$ of the past of an event $r \succ p, q$ is no longer as simple as Eqn.~(\ref{voldef}). To begin with,  every $A$ lies in an uncountable infinity of spatial hypersurfaces $\{\Sigma\}$ and the  choice of $\Sigma$ determines $V$. In particular  the factor replacing $\zeta_\dm$ in Eqn.~(\ref{voldef}) will depend on the induced spatial geometry on $\Sigma$ as well as the location of $p, q \in \Sigma$. This information can then be used to calculate the discrete distance $\dd$ using 
Eqn.~(\ref{delta}). Again, randomness ensures that $\dd > \cdpp$, which translates into asymptotic silence.

To prove this more rigorously requires  more extensive arguments, along the lines of \cite{Major:2006hv,Major:2009cw}. 
In this work we will resort to the easier task of performing numerical simulations to support this claim. We  obtain causal sets from a Poisson sprinkling  into regions of $\dm=2,3$ and $4$ dimensional Minkowski spacetime. Because of the calculational complications that arise in considering an arbitrary antichain $A$ we pick the region to be a $\dm$ dimensional ``box'' or hypercube, with the  intial and final hypersurfaces corresponding to $t=0$ and $t=L$, respectively.  Moreover, we choose $A$ to be the past-most inextendible antichain, so that it  is reasonably well approximated by  the $t=0$ slice. This then limits the systematic errors in the calculation. These errors come from our use of $\zeta_\dm$ as defined in Eqn.~(\ref{zeta}) even though the elements in $A$ do  not strictly lie in the $t=0$ slice. By doing so, $\dd$ is sometimes a little \textsl{underestimated}.
Defining 
\be
\Delta \equiv \frac{\delta-\cdpp}{\cdpp},
\ee
as a measure of asymptotic silence, we see that 
it becomes slightly negative { for some datapoints} even for large distances when there is better agreement with the  continuum.

\begin{figure}[!t]
\includegraphics[width=0.45\linewidth]{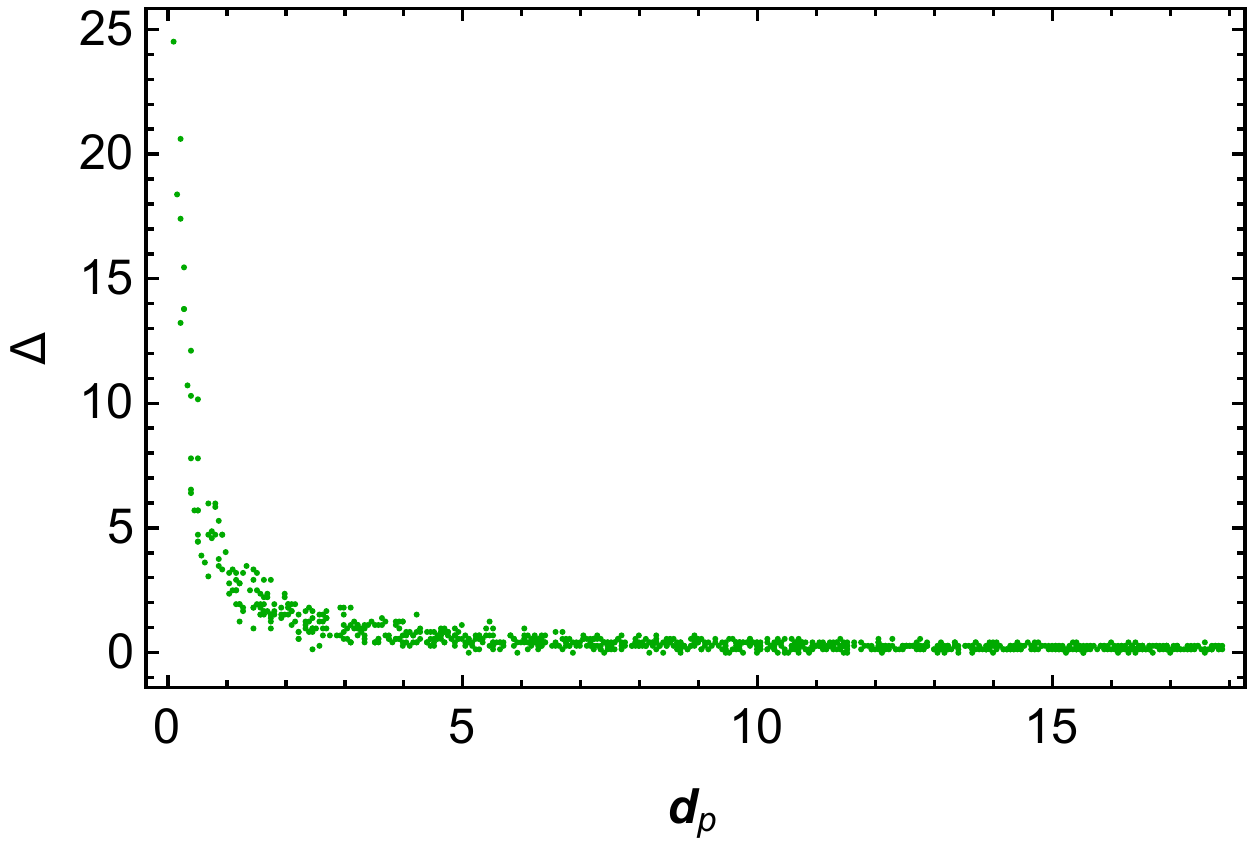}\quad\includegraphics[width=0.45\linewidth]{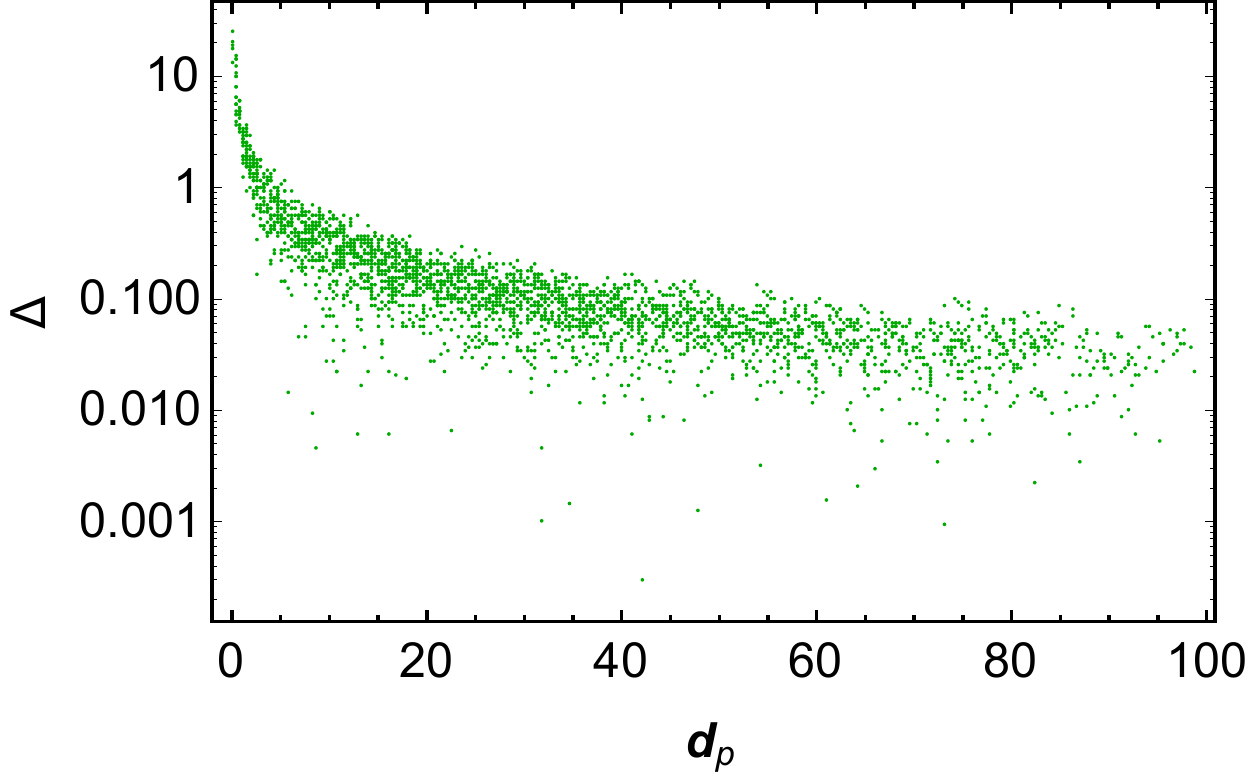}
\caption{\label{fig:2d_AS} We show $\Delta$ as a function of the dimensionless spatial distance $\cdpp$ for $10^5$ elements for a causal set approximated by $1+1$ dimensional Minkowski spacetime, and for $100$ trials. The logarithmic plot on the right shows  the slow approach to the exact infrared limit, whereas the linear plot highlights that asymptotic silence does not persist beyond $\cdpp \approx 10$.}
\end{figure}

The simulation is performed by picking an element  $e\in A$ at random and finding its  causal set distance $\dd(e)$ to all other elements  in $A$. The proper distance $\cdpp$ can simultaneously be calculated from the embedding coordinates  of $C$ into $\dm$ dimensional  Minkowski spacetime.   Ideally a good comparison is possible only if the length of the sprinkling box  is the same in all dimensions. Starting with a box of size $100^2$ in $\dm=2$ this translates into  $N=10^6$ and $N=10^8$ in $\dm=3,4$. These however need  $64$ and $ 6400$ GBs of RAM, respectively\footnote{For a causal set the number of possible relations is $\binom{N}{2}$, which tranlates into $\sim N^2/16$ GB of RAM.}. While the former is possible to achieve on modern machines, the latter is  not. We will therefore content ourselves with the requirement that the box is large enough to begin to see emergent continuum behaviour. Our simulations are done for  $N=10,000$ elements in $\dm=2$, $N=100,000$ elements for $\dm=3$ and $N=400,000$ elements for $\dm=4$. Because of the ease of simulation in $\dm=2$, we perform 100 trials which gives us independent data that can be plotted on the same graph as shown in Figure \ref{fig:2d_AS}. Our simulations were done using the Causal Set Cactus package \cite{cactustoolkit}. 

\begin{figure}[!t]
\includegraphics[width=0.45\linewidth]{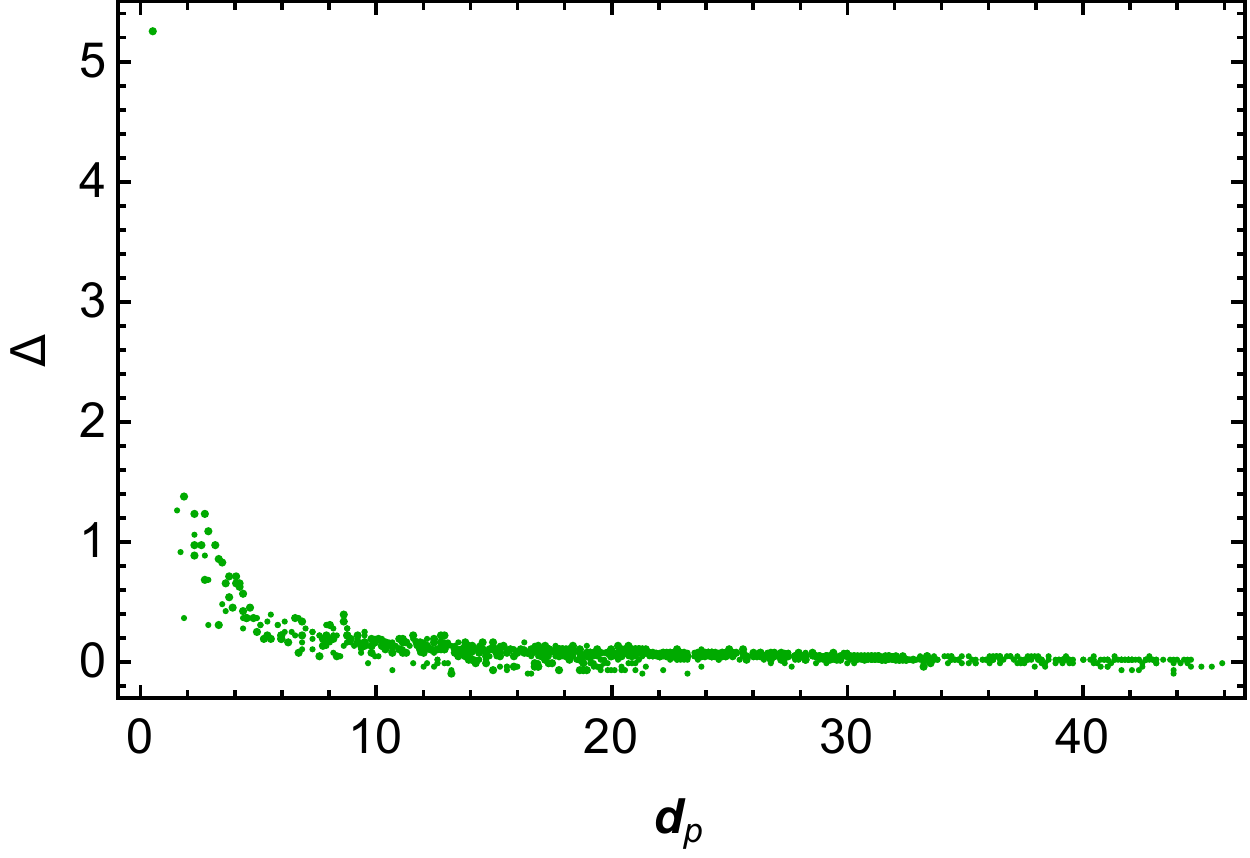}\quad\includegraphics[width=0.45\linewidth]{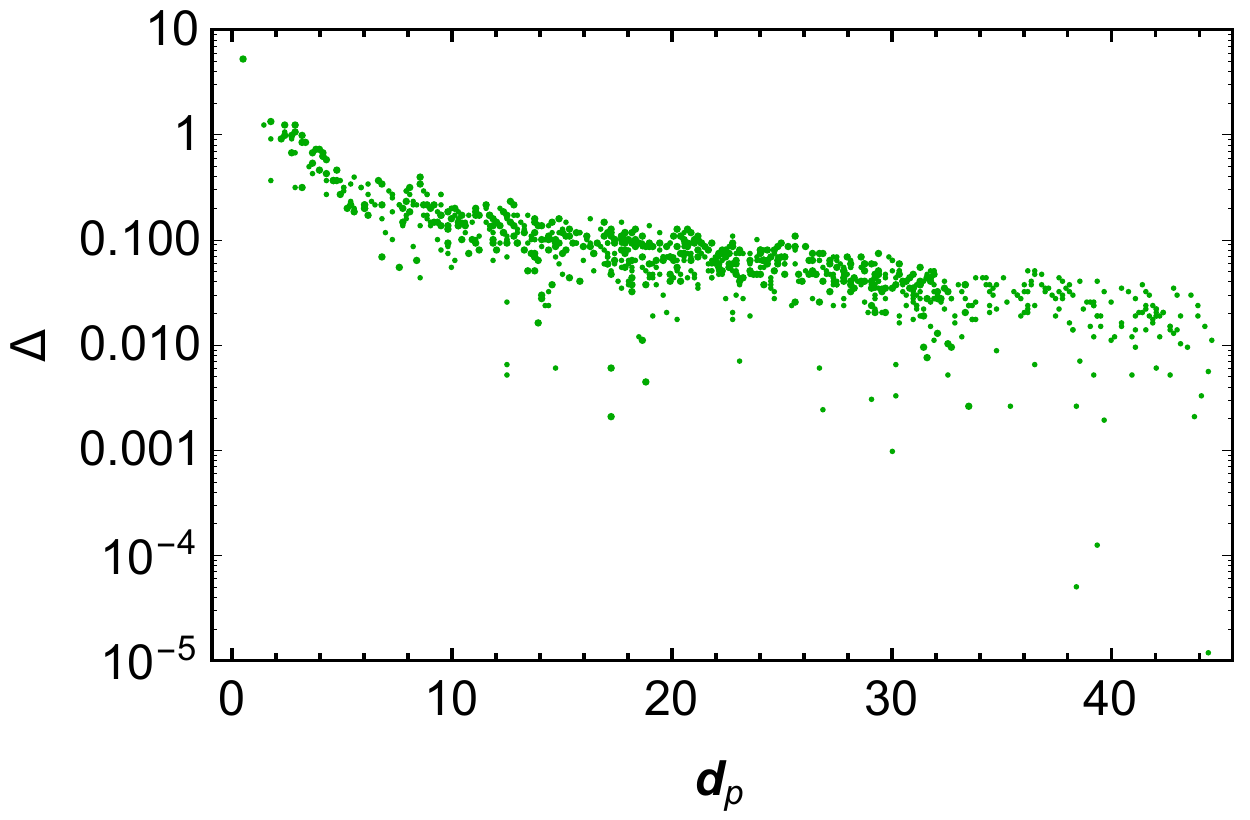}
\caption{\label{fig:3d_AS} We show $\Delta$ as a function of the dimensionless spatial distance $\cdpp$ for a $10^6$ element causal set  approximated by  $2+1$ dimensional Minkowski spacetime.}
\end{figure}

\begin{figure}[!t]
\includegraphics[width=0.45\linewidth]{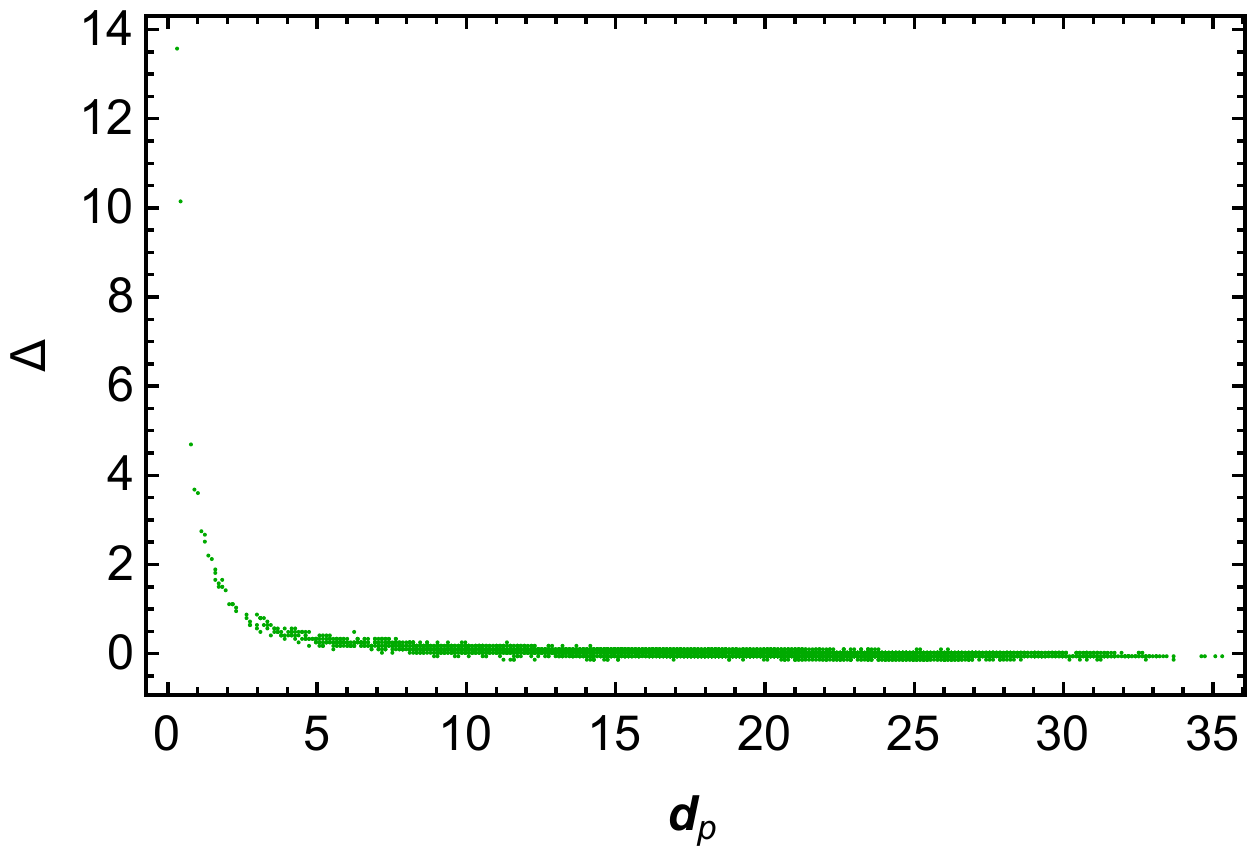}\quad\includegraphics[width=0.45\linewidth]{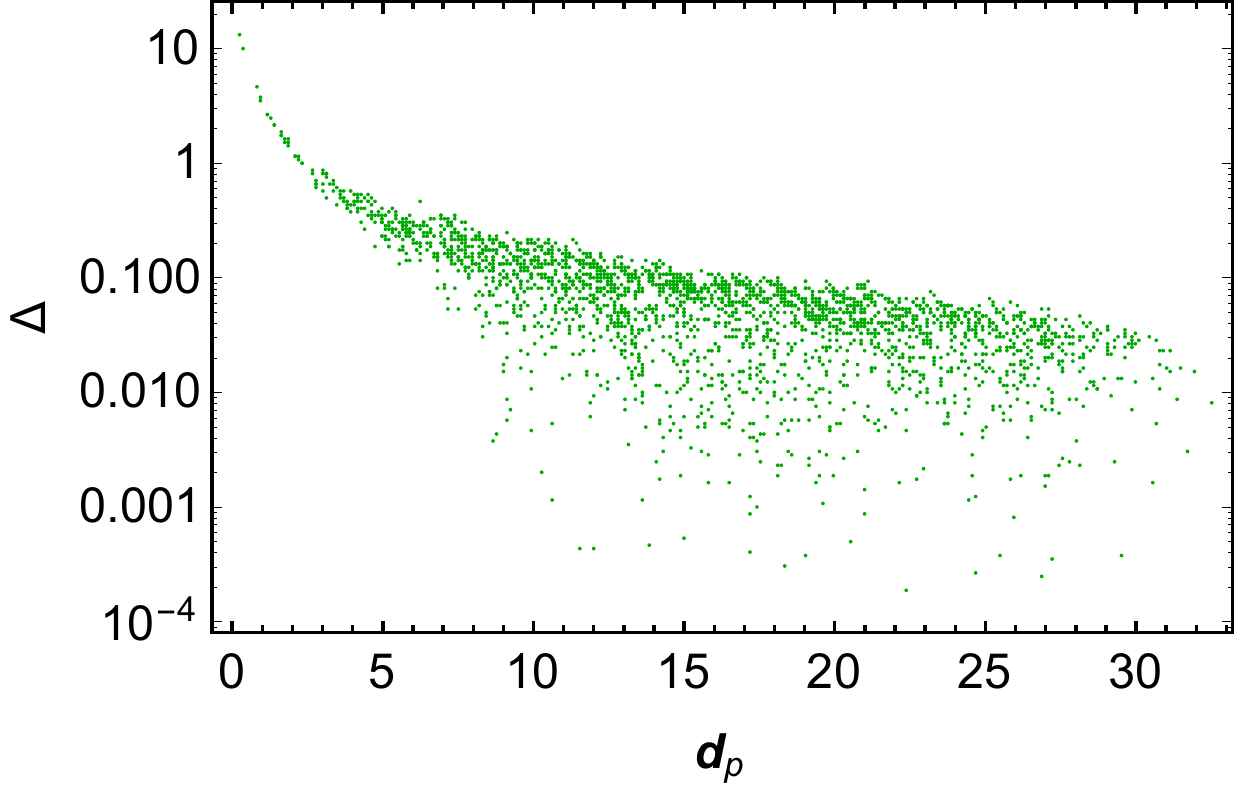}
\caption{\label{fig:4d_AS} We show $\Delta$ as a function of the dimensionless spatial distance $\cdpp$ for  a $4 \cdot 10^6$ element element causal set  approximated by $3+1$ dimensional Minkowski spacetime.}
\end{figure}

In Figs.~\ref{fig:2d_AS}, \ref{fig:3d_AS} and \ref{fig:4d_AS}, we see that  $\Delta$ drops sharply down to near zero at a continuum distance $\cdp \sim 10\,  l_c$ in all cases. Thus, as it should be, the causal set is well-approximated by the continuum manifold 
{at scales larger than $10\, l_c$}, beyond which the discrete and the continuum distance merge. Around $\cdp \simeq 10\, l_c$, the approach of $\Delta$ to zero slows down significantly; thus small remnants of the discreteness remain. 
For smaller $\cdpp$, 
$\Delta$ becomes fairly  large, in particular even significantly larger than one, thus supporting the hypothesis of asymptotic silence.

\begin{figure}[!t]
\begin{center}
	\includegraphics[width=0.3 \linewidth,clip=true,trim=11cm 8cm 12cm 5cm]{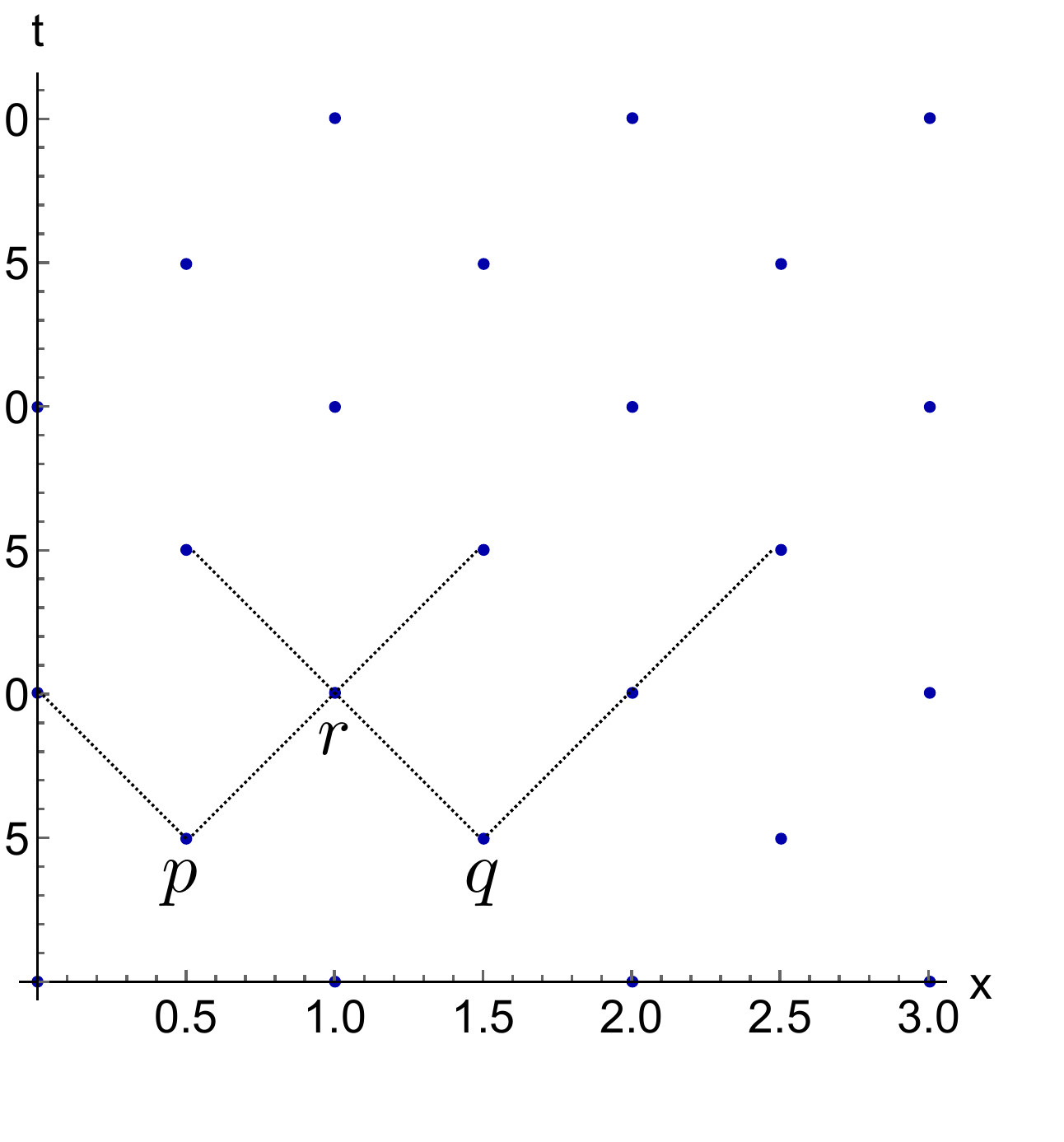}
	\caption{\label{lattice_illustration} We show a regular lightcone lattice in 1+1 dimensions. The points $p,q$ and $r$ to illustrate that asymptotic silence does not occur.}
\end{center}
\end{figure}

To distinguish causal set discreteness from that of a regular lattice, it  is instructive to consider a lightcone lattice in two dimensions, cf.~Fig.~\ref{lattice_illustration}. 
In contrast with the causal set case, the lattice breaks Lorentz invariance by  introducing a preferred time foliation. Because of this, for every pair of points $p$ and $q$ on the same time slice, there is a unique point $r$, which lies exactly on the intersection of the lightcones of $p$ and $q$, cf.~Fig.~\ref{lattice_illustration}. Even if the lattice were tailored so that $\dd>\cdpp$, the offset would be constant with $\cdpp$.
Hence, calculating the distance measure $\tilde{\delta}$ gives the same answer as in the continuum. We conclude that a deviation of $\dd$ from $\cdpp$ is not a generic artifact of discretisation. Instead we attribute the observation of $\Delta>0$ to the randomness associated with causal set 
discreteness.

\begin{figure}[!t]
\begin{center}
\includegraphics[width=0.5\linewidth]{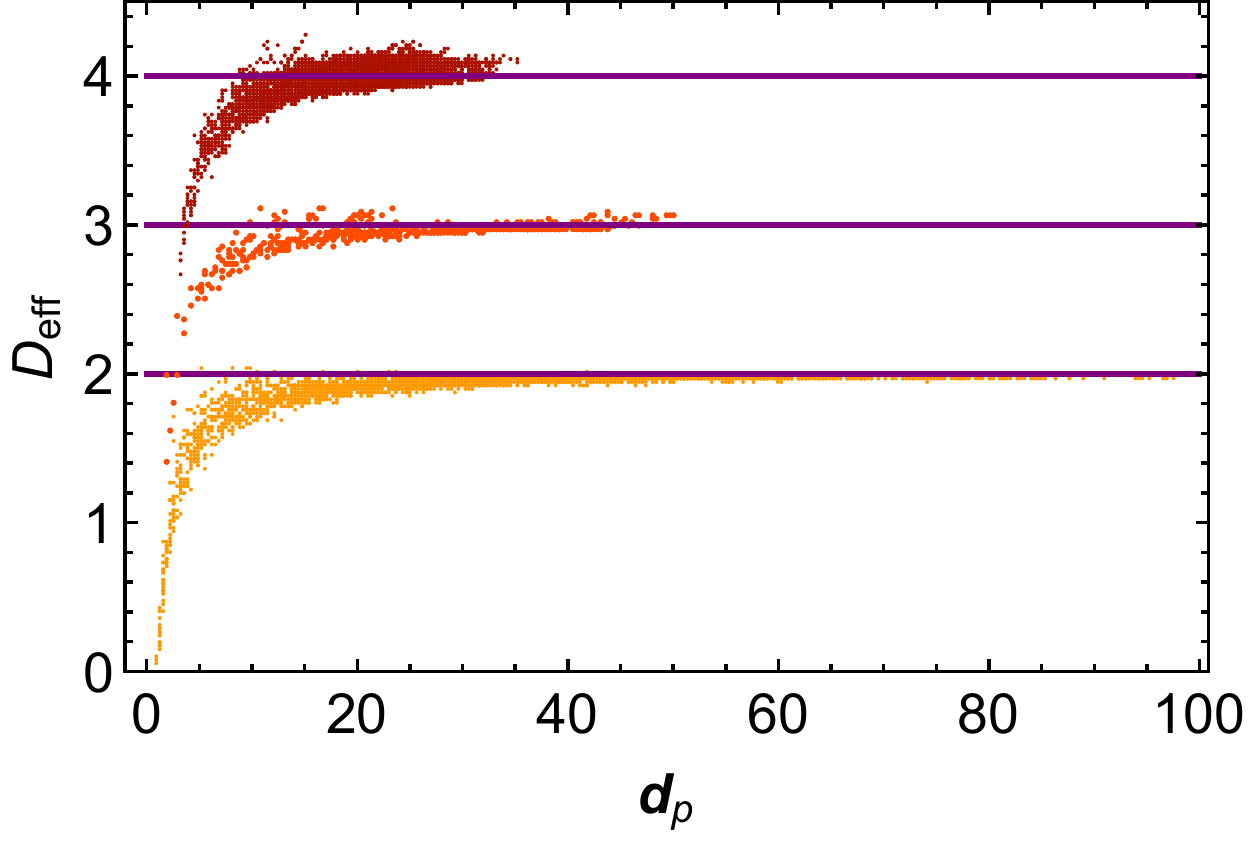}
\end{center}
\caption{\label{fig:dim_red} We show the effective dimension $D_{\rm eff}$ as a function of  $\cdpp$ for causal sets approximated by   $D=2$ (orange dots), $D=3$ (red dots) and $D=4$ (dark red dots) Minkowski spacetimes. Each case exhibits dimensional reduction at small scales and approaches the continuum dimension at large scales. We have excluded points with $\cdpp <l_c$.}
\end{figure}

We can now 
extract an effective dimension $\dmeff$ by comparing the volumes arising from $\dd$ and $\cdpp$,  
\be \left( \dd\right)^\dmeff = \left(\cdpp\right)^{\dm}.  \ee 
The right hand side of the expression is the continuum spacetime volume of a box with sides of length $\dd$ in terms of the cut-off scale $V_c$.  $\dmeff$ is thus defined  by asking what the dimension of a hypercube of side $\dd$ {\it should} be if it is to give the same continuum volume as that obtained from $\cdpp$.  This comparison of volumes is only one way of defining the effective dimension, but it does lead to a reduced dimension 
\be \dmeff \equiv \dm \, \frac{\ln (\cdpp)}{\ln(\dd)} < \dm.  
\ee 
As $\dd$ approaches $\cdpp$, $\dmeff$ goes over to $\dm$, but for small $\cdpp$ it is less than $\dm$, cf.~Fig.~\ref{fig:dim_red}. Indeed 
it becomes negative for $\cdp < l_c$  (note that values $\cdp<l_c$ are not included in the plot). This may appear surprising, but is a result of the randomness. Namely if $L=(\frac{V}{V_c})^\frac{1}{\dm}$ is the length of the box in units of $l_c$, $\cdpp \in (0, L)$ and  cannot  always be greater than $l_c$; indeed there is a small but non-vanishing probability that it is smaller. We must therefore view $\dmeff$ more as a sign of dimensional reduction rather than interpret its value literally in the ultraviolet. Thus, dimensional reduction, present in many other approaches to quantum gravity,
can in this sense also be seen in causal sets\footnote{Note that this is not in contradiction to the \emph{increase} of the \emph{spectral} dimension in causal sets \cite{Eichhorn:2013ova}, as different dimensional estimators can exhibit different behaviour in the quantum gravity regime (see, e.g., \cite{Reuter:2011ah}). }. 
Fig.~\ref{fig:dim_red} shows the behaviour of this effective dimension for $\dm=2,3$ and $4$.   A  fairly striking feature of this figure is that the convergence to the continuum occurs at $\cdp\sim 10  l_c$ for all dimensions, which marks the scale of kinematic discretisation. 

We conclude by bringing attention to another property of causal sets that could be connected to asymptotic silence\footnote{This idea is due to Rafael Sorkin.}. Specifically, in an asymptotically silent spacetime, neighbouring worldlines do not meet, and so it becomes impossible for particles to scatter off each other. In causal sets, one might expect to observe a similar property for particle scattering at high center-of-mass-energies: A scattering event at center-of-mass-energy $M$ takes up a spacetime volume that decreases with increasing $M$. Accordingly, scattering events at  center-of-mass-energies corresponding to the inverse cut-off scale should be expected to take place within a spacetime-interaction region of volume $V_c$. In a causal set, the probability that there is a causal set element within a given volume rapidly shrinks to zero as one reaches the cut-off scale. Therefore, in a causal set universe all particles become ``transparent" to each other at very high center-of-mass energies, and scattering no longer takes place. It is interesting to observe that another approach to quantum gravity, namely the asymptotic safety scenario, exhibits hints of a similar behavior at high energies \cite{Litim:2007iu,Dobrich:2012nv}, and in fact also exhibits signs of dimensional reduction which have been discussed in relation to asymptotic silence.\newline

\begin{acknowledgments} We thank Rafael Sorkin for discussions. We also thank David Rideout for help with Cactus. The simulations were performed using the Cactus Causal Set Toolkit \cite{cactustoolkit}.  A.~E.~is supported by the DFG under the Emmy-Noether program, grant no.~EI-1037-1, and by an Emmy-Noether visiting fellowship at the Perimeter Institute for Theoretical Physics. Research at Perimeter Institute is supported by the Government of Canada through Industry Canada and by the Province of Ontario through the Ministry of Economic Development \& Innovation. S.~S.~was supported in part under an agreement with Theiss Research and funded by a grant from the FQXI Fund on the basis of proposal FQXi-RFP3-1346 to the Foundational Questions Institute.  \end{acknowledgments}

\end{document}